\documentclass[12pt,english]{elsarticle}
\usepackage[T1]{fontenc}
\usepackage[latin9]{inputenc}
\usepackage{amsbsy}
\usepackage{amstext}
\usepackage{amssymb}
\usepackage{graphicx}
\usepackage{esint}

\makeatletter

\providecommand{\tabularnewline}{\\}

\journal{PLB}

\makeatother

\usepackage{babel}
\begin{document}

\title{Quantum gravity stability of isotropy in homogeneous cosmology}

\author{Bogus\l{}aw Broda}

\ead{bobroda@uni.lodz.pl}

\ead{http://merlin.fic.uni.lodz.pl/kft/people/BBroda}

\address{Department of Theoretical Physics, University of \L\'od\'{z}, Pomorska
149/153, PL--90-236~\L\'od\'{z}, Poland}
\begin{abstract}
It has been shown that anisotropy of homogeneous spacetime described
by the general Kasner metric can be damped by quantum fluctuations
coming from perturbative quantum gravity in one-loop approximation.
Also, a formal argument, not limited to one-loop approximation, is
put forward in favor of stability of isotropy in the exactly isotropic
case.\end{abstract}
\maketitle
\def\citet{\cite}

\section{Introduction}

Standard \citet{Jerusalem} and loop \citet{ABL,Boj,AS} quantum cosmology
heavily depends on the implicit assumption of (quantum) stability
of general form of the metric. As a principal starting point in quantum
cosmology, one usually chooses a metric of a particular (more or less
symmetric) form. In the simplest, homogeneous and isotropic case,
the metric chosen is the (flat) Friedmann--Lema�tre--Robertson--Walker
(FLRW) one. Consequently, (field theory) quantum gravity reduces to
a much more tractable quantum mechanical system with a finite number
of degrees of freedom. It is obvious that such an approach greatly
simplifies quantum analysis of cosmological evolution, but under no
circumstances is it obvious to what extent is such an approach reliable.
The quantum cosmology approach could be considered unreliable when
(for example) the assumed symmetry of the metric would be unstable
due to quantum fluctuations. More precisely, in the context of the
stability, one can put forward the two, to some extent complementary,
issues (questions): (1) assuming a small anisotropy in the almost
isotropic cosmological model, have quantum fluctuations a tendency
to increase the anisotropy or, just the opposite, to reduce it? (2)
assuming we start quantum evolution from an exactly isotropic metric
should be we sure that no quantum fluctuations are able to perturb
the isotropy?

In this Letter, we are going to address the both issues of the quantum
stability of spacetime metric in the framework of standard covariant
quantum gravity. Namely, in Section 2, we address the first stability
issue for an anisotropic (homogeneous) metric of the Kasner type,
to one loop in perturbative expansion. In Section 3, we give a simple,
formal argument, not limited to one loop, concerning the second issue.

\section{One-loop stability}

The approach applied in this section is a generalization of our approach
used in \citet{Bro} in the context of FLRW geometry. In our present
work, the starting point is an anisotropic (homogeneous) metric,

\begin{equation}
ds^{2}=dt^{2}-a_{1}^{2}(t)\left(dx^{1}\right)^{2}-a_{2}^{2}(t)\left(dx^{2}\right)^{2}-a_{3}^{2}(t)\left(dx^{3}\right)^{2},\label{eq:ds2-ogolnametryka}
\end{equation}
 of the Kasner type, i.e.

\begin{equation}
a_{i}^{2}(t)\equiv\left|\frac{t}{t_{0}}\right|^{2k_{i}},\qquad i=1,2,3,\label{eq:a2-czynnikiskali}
\end{equation}
where $k_{i}$ are the Kasner exponents. One should stress that we
ignore any assumptions concerning matter content, and consequently,
no prior bounds are imposed on $k_{i}$. 

In the perturbative approach
\begin{equation}
g_{\mu\nu}=\eta_{\mu\nu}+\kappa h_{\mu\nu},\label{eq:g-minkowskiplush}
\end{equation}
then
\begin{equation}
\kappa h_{i}(t)=1-\left|\frac{t}{t_{0}}\right|^{2k_{i}},\qquad h_{i}(t_{0})=0,\label{eq:a2-przezh}
\end{equation}
where $\kappa=\sqrt{32\pi G_{N}}$, with $G_{N}$---the Newton gravitational
constant. The quantized field 
\begin{equation}
h_{\mu\nu}(t,\boldsymbol{x})=\left(\begin{array}{cccc}
0 & 0 & 0 & 0\\
0 & h_{1}(t) & 0 & 0\\
0 & 0 & h_{2}(t) & 0\\
0 & 0 & 0 & h_{3}(t)
\end{array}\right)\equiv\textrm{diag}\left(0,h_{i}\left(t\right)\right)\label{eq:h-diagh}
\end{equation}
is small, as expected, closely to the expansion (reference) point
$t_{0}$. Using the gauge freedom to satisfy the harmonic gauge condition
(see, the second formula in $\left(\ref{eq:hbardiag}\right)$), we
gauge transform the gravitational field $h_{\mu\nu}$ as follows,
\begin{equation}
h_{\mu\nu}\rightarrow h_{\mu\nu}^{'}=h_{\mu\nu}+\partial_{\mu}\xi_{\nu}+\partial_{\nu}\xi_{\mu},\label{eq:h-hprim}
\end{equation}
where the gauge parameter
\begin{equation}
\xi_{\mu}\left(t\right)=\left(-\frac{1}{2}\int_{0}^{t}h(t')\, dt',\,0,\,0,\,0\right),\qquad h\left(t\right)\equiv h_{1}\left(t\right)+h_{2}\left(t\right)+h_{3}\left(t\right).\label{eq:ksi}
\end{equation}
Then,

\begin{equation}
h_{\mu\nu}^{'}(t,\boldsymbol{x})=\textrm{diag}\left(2\dot{\xi}_{0}\left(t\right),h_{i}\left(t\right)\right)=\textrm{diag}\left(-h\left(t\right),h_{i}\left(t\right)\right),\label{eq:hprim}
\end{equation}
and, skipping the prime for simplicity, we have
\begin{equation}
h{}_{\lambda}^{\lambda}(t)=-2h(t),\label{eq:sladh}
\end{equation}
where spacetime indices are being manipulated with the Minkowski metric
$\eta_{\mu\nu}$. Now, we should switch from our present $h_{\mu\nu}$
to standard perturbative gravitational variables, i.e.\ to the {}``barred''
field $\bar{h}_{\mu\nu}$ defined by
\begin{equation}
\bar{h}_{\mu\nu}\equiv h_{\mu\nu}-\frac{1}{2}\eta_{\mu\nu}h_{\lambda}^{\lambda},\label{eq:habar}
\end{equation}
and 
\begin{equation}
\bar{h}_{\mu\nu}(t,\boldsymbol{x})=\textrm{diag}\left(0,h_{i}\left(t\right)-h(t)\right)\quad\textrm{with}\quad\partial^{\mu}\bar{h}_{\mu\nu}=0.\label{eq:hbardiag}
\end{equation}
The Fourier transform of $\bar{h}_{\mu\nu}$ is
\begin{equation}
\tilde{\bar{h}}_{\mu\nu}(p)\equiv\tilde{\bar{h}}_{\mu\nu}(\omega,\boldsymbol{p})=\left(2\pi\right)^{3}\delta^{3}(\boldsymbol{p})\,\textrm{diag}\left(0,\tilde{h_{i}}(\omega)-\tilde{h}(\omega)\right),\label{eq:fourierhabar}
\end{equation}
where, for the $h_{i}$ of the explicit form $\left(\ref{eq:a2-przezh}\right)$,
we have (from now on, we denote classical gravitational fields with
the superscript {}``c'') 
\begin{equation}
\kappa\tilde{h_{i}^{\mathrm{c}}}(\omega)=2\pi\delta\left(\omega\right)+2t_{0}^{-2k_{i}}\sin\left(\pi k_{i}\right)\Gamma\left(2k_{i}+1\right)\left|\omega\right|^{-2k_{i}-1}.\label{eq:hcomegafourier}
\end{equation}
According to $\left(\ref{eq:hqprzezhcfinal}\right)$ a one-loop quantum
contribution corresponding to the classical metric $\left(\ref{eq:fourierhabar}\right)$
equals

\begin{eqnarray}
\tilde{h^{\textrm{q}}}_{\mu\nu}(p) & = & \frac{\pi\kappa^{2}p^{2}}{2}\log\left(\frac{-p^{2}}{\Lambda^{2}}\right)\delta^{3}(\boldsymbol{p})\left[(2\alpha_{1}\mathbb{E}+4\alpha_{2}\mathbb{P})\,\textrm{diag}\left(0,\tilde{h_{i}^{\mathrm{c}}}-\tilde{h^{\mathrm{c}}}\right)\right]_{\mu\nu}\nonumber \\
 & = & 2\pi\kappa^{2}\omega^{2}\log\left|\frac{\omega}{\Lambda}\right|\delta^{3}(\boldsymbol{p})\,\textrm{diag}\left(\alpha_{2}\tilde{h^{\mathrm{c}}},\alpha_{1}\tilde{h_{i}^{\mathrm{c}}}-\left(\alpha_{1}+\alpha_{2}\right)\tilde{h^{\mathrm{c}}}\right).\label{eq:hqprzezdiaghc}
\end{eqnarray}
Defining the auxiliary function
\begin{equation}
\tilde{h_{i}^{\mathrm{\textrm{Q}}}}(\omega)\equiv\omega^{2}\log\left|\frac{\omega}{\Lambda}\right|\tilde{h_{i}^{\mathrm{c}}}(\omega),\label{eq:hQdef}
\end{equation}
we have
\begin{eqnarray}
\kappa\tilde{h_{i}^{\mathrm{\textrm{Q}}}}(\omega) & = & 2\pi\delta\left(\omega\right)\omega^{2}\log\left|\frac{\omega}{\Lambda}\right|\nonumber \\
 & + & 2t_{0}^{-2k_{i}}\sin\left(\pi k_{i}\right)\Gamma\left(2k_{i}+1\right)\left|\omega\right|^{-2k_{i}+1}\log\left|\frac{\omega}{\Lambda}\right|.\label{eq:hQjawnie}
\end{eqnarray}
\begin{figure}
\begin{centering}
\includegraphics[angle=270,scale=0.3]{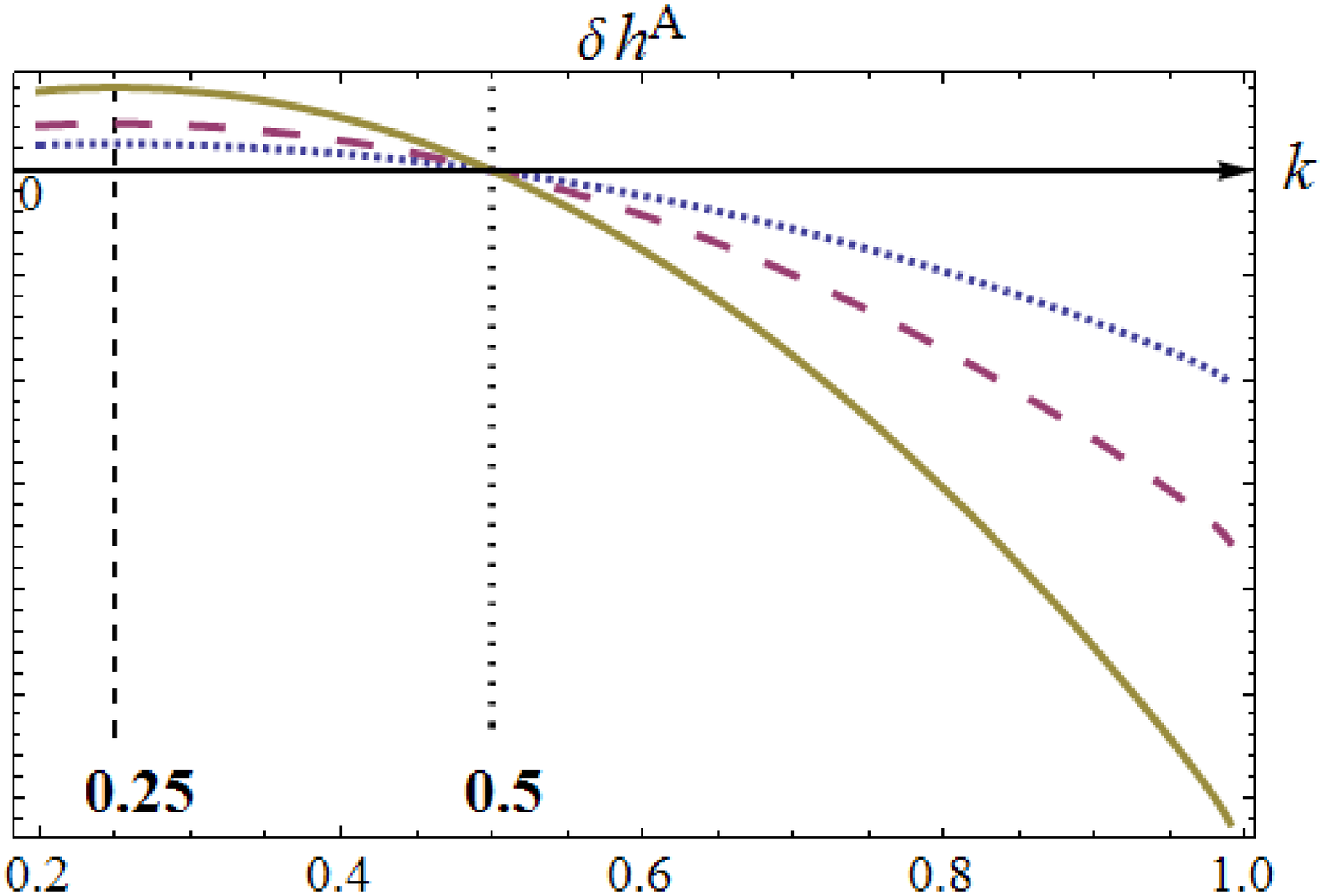}\caption{\label{fig:Rysunek}}

\par\end{centering}

\noindent \centering{}{\small The drawing qualitatively presents 3
demonstration curves (for 3 different UV cutoffs $\Lambda$) of the
function $\delta h^{\textrm{A}}\left(k\right)$ defined by $\left(\ref{eq:anisotropic}\right)$
(and $\left(\ref{eq:hQtjawnie}\right)$). For the Kasner exponents
$k\in\left(\frac{1}{4},1\right)$, $\delta h^{\textrm{A}}\left(k\right)$
is evidently decreasing function thus supporting the damping of cosmological
anisotropy. }
\end{figure}
Its Fourier reverse is
\begin{eqnarray}
\kappa h_{i}^{\textrm{Q}}(t) & = & 2t_{0}^{-2k_{i}}k_{i}\left(2k_{i}-1\right)\left|t\right|^{2k_{i}-2}\nonumber \\
 & \times & \left[\psi\left(2-2k_{i}\right)+\frac{\pi}{2}\tan\left(\pi k_{i}\right)-\log\left|\Lambda t\right|\right],\label{eq:hQtjawnie}
\end{eqnarray}
where $\psi$ is the digamma function, and according to $\left(\ref{eq:hqprzezdiaghc}\right)$
\begin{equation}
h_{\mu\nu}^{\textrm{q}}(t,\boldsymbol{x})=\left(\frac{\kappa}{2\pi}\right)^{2}\textrm{diag}\left(\alpha_{2}h^{\textrm{Q}}(t),\alpha_{1}h_{i}^{\textrm{Q}}(t)-\left(\alpha_{1}+\alpha_{2}\right)h^{\textrm{Q}}(t)\right).\label{eq:hqtxjawnie}
\end{equation}
Performing the gauge transformation in the spirit of $\left(\ref{eq:ksi}\right)$
we can remove the first (time) component in $\left(\ref{eq:hqtxjawnie}\right)$,
and (once more, skipping the prime for simplicity) we get a quantum
contribution to the Kasner metric 
\begin{equation}
h_{\mu\nu}^{\textrm{q}}(t,\boldsymbol{x})=\left(\frac{\kappa}{2\pi}\right)^{2}\textrm{diag}\left(0,\alpha_{1}h_{i}^{\textrm{Q}}(t)-\left(\alpha_{1}+\alpha_{2}\right)h^{\textrm{Q}}(t)\right).\label{eq:hqtxprim}
\end{equation}
Only the {}``anisotropic'' part of $\left(\ref{eq:hqtxprim}\right)$,
i.e.
\begin{equation}
\delta h_{i}^{\textrm{A}}=\left(\frac{\kappa}{2\pi}\right)^{2}\alpha_{1}h_{i}^{\textrm{Q}},\label{eq:anisotropic}
\end{equation}
can influence the anisotropy of the evolution of the Universe. Since
the dependence of $\delta h_{i}^{\textrm{A}}$ on $k_{j}$ is purely
{}``diagonal'' ($\delta h_{i}^{\textrm{A}}$ depends only on $k_{j}$
with $j=i$, see $\left(\ref{eq:hQtjawnie}\right)$), we have the
following simple rule governing (de)stabilization of the isotropy:
the increasing function $\delta h^{\textrm{A}}(k)$ implies destabilization
(there is a greater contribution of quantum origin to the metric in
the direction of a greater classical expansion), whereas the decreasing
function implies stabilization. Unfortunately, $\delta h^{\textrm{A}}\left(k\right)$
is not a monotonic function because the digamma function $\psi$ oscillates,
and moreover $\left(\ref{eq:hQtjawnie}\right)$ is (in general%
\footnote{The quantum contribution is $\Lambda$-cutoff independent for $k_{i}=\frac{1}{2}$
(a limit in $\left(\ref{eq:hQtjawnie}\right)$ exists), i.e.\ for
pure radiation (see, \citet{Bro}). Intuitively, it could be explained
by the fact that a scale-independent classical source, the photon
field, implies vanishing of scale-dependent logarithms (no quantum
{}``anomaly'').%
}) a $\Lambda$-cutoff dependent function. Nevertheless, if we assume
the point of view that it is not necessary to expect or require the
stability of the isotropy in the whole domain of the Kasner exponents
$k_{i}$, but only for some subset of them, considered physically
preferred, a definite answer emerges. Since $k=\frac{1}{2}$ corresponds
to radiation, and $k=\frac{2}{3}$ corresponds to matter, we could
be fully satisfied knowing that $\delta h^{\textrm{A}}\left(k\right)$
is monotonic in the interval $k\in\left(\frac{1}{4},1\right)$($\supset\left[\frac{1}{2},\frac{2}{3}\right]$).
Furthermore, since $\alpha_{1}>0$ for any spin (see, Table \ref{tab:tabela}),
$\delta h^{\textrm{A}}\left(k\right)$ is a decreasing function in
this interval, implying (quantum) damping of the anisotropy (see,
Fig. \ref{fig:Rysunek}).

\section{Above one loop and final remarks}

Section 2 has been limited to one-loop perturbative analysis of the
stability of isotropy of cosmological evolution. But one can give
a simple, formal argument ensuring stability of the {}``exactly isotropic''
expansion, i.e. for 
\begin{equation}
k_{1}=k_{2}=k_{3},\label{eq:kirowne}
\end{equation}
which is perturbative but not limited to one loop, making use of $\left(\ref{eq:hqprzezhc}\right)$,
where now $D$ and $\Pi$ is the full propagator and the full vacuum
polarization, respectively. Since
\begin{equation}
h_{\mu\nu}^{\textrm{c}}=\textrm{diag}\left(0,h\left(t\right),h\left(t\right),h\left(t\right)\right),\label{eq:diagh}
\end{equation}
no spatial coordinate $x_{i}$ is singled out in $\left(\ref{eq:diagh}\right)$,
and consequently no spatial coordinate can be singled out on LHS of
$\left(\ref{eq:hqprzezhc}\right)$. This argument is only of purely
formal interest as any other fluctuations can destabilize the isotropy.

Recapitulating, as far as perturbative quantum gravity in one-loop
approximation is concerned we have observed that in a (hopefully)
physically preferred region of the Kasner exponents $k_{i}\in\left(\frac{1}{4},1\right)$
we should expect damping of anisotropy by quantum fluctuations, thus
supporting reliability of the approach of quantum cosmology in this
regime. One should point out that this result is subjected to several
limitations. First of all, since $\kappa h_{i}(t)$ should be small,
$t$ should be close to $t_{0}$ because of $\left(\ref{eq:a2-przezh}\right)$.
But for $t\thickapprox t_{0}$, by virtue of $\left(\ref{eq:hqtxjawnie}\right)$,
we have
\begin{equation}
h_{ii}^{\textrm{q}}\thicksim\kappa^{2}t_{0}^{-2k}\left|t\right|^{2k-2}\thicksim\kappa^{2}t_{0}^{-2}\sim\left(\frac{t_{\textrm{Planck}}}{t_{0}}\right)^{2}.\label{eq:orderh}
\end{equation}
Therefore, to stay in the perturbative regime, $t_{0}$ should be
greater than $t_{\textrm{Planck}}$, and we have to be away from the
(primordial) classical singularity. Instead, for $t_{0}$ many orders
greater than $t_{\textrm{Planck}}$, according to $\left(\ref{eq:orderh}\right)$
the quantum contribution becomes small, and moreover classical matter
(including radiation) is expected to begin to play a role. In particular,
it was shown in \citet{Mis}%
\footnote{The author would like to thank the Referee for pointing out the reference.%
} that the effects of viscosity in the radiation and pressure from
collisionless radiation ensure isotropization (and stability) of cosmological
evolution at late times.

\section*{Acknowledgements}

I would like to thank prof.~Ashbay Ashtekar for his e-mail with clarifications
concerning my previous paper \citet{Bro}, and for bringing my attention
to the reference \citet{AS}.

\appendix

\section{One-loop vacuum polarization}

For the Reader's convenience, we present here a short derivation of
the one-loop quantum correction to a classical gravitational field
which is coming from (one-loop) vacuum polarization (self-energy)
(for more details, see, \citet{Bro}, and also compare to \citet{Duf,Don}
for $\boldsymbol{x}$-dependent but $t$-independent case).

\begin{table}
\begin{centering}
\begin{tabular}{cccccc}
\hline 
\noalign{\vskip\doublerulesep}
 & spin &  & $\alpha_{1}$ & $\alpha_{2}$ & $\alpha_{1}+\alpha_{2}$\tabularnewline[\doublerulesep]
\noalign{\vskip\doublerulesep}
\hline 
\noalign{\vskip\doublerulesep}
 & 0 &  & $\frac{1}{480}$ & -$\frac{1}{720}$ & $\frac{1}{1440}$\tabularnewline[\doublerulesep]
\noalign{\vskip\doublerulesep}
\noalign{\vskip\doublerulesep}
 & $\frac{1}{2}$ &  & $\frac{1}{160}$ & -$\frac{1}{240}$ & $\frac{1}{480}$\tabularnewline[\doublerulesep]
\noalign{\vskip\doublerulesep}
\noalign{\vskip\doublerulesep}
 & 1 &  & $\frac{1}{40}$ & -$\frac{1}{60}$ & $\frac{1}{120}$\tabularnewline[\doublerulesep]
\noalign{\vskip\doublerulesep}
\noalign{\vskip\doublerulesep}
 & 2 &  & $\frac{27}{80}$ & -$\frac{59}{240}$ & $\frac{11}{120}$\tabularnewline[\doublerulesep]
\hline 
\noalign{\vskip\doublerulesep}
\end{tabular}\caption{\label{tab:tabela}}

\par\end{centering}

\noindent \centering{}{\small The coefficients entering $\left(\ref{eq:hqprzezhcfinal}\right)$
taken from \citet{CD,CLM,Cap,CDH} (see, also \citet{DL}. In particular,
$\alpha_{1}$ enters $\left(\ref{eq:anisotropic}\right)$, and its
positivity for any spin supports the damping of cosmological anisotropy. }
\end{table}

In the momentum representation, the lowest order quantum corrections
$\tilde{\bar{h^{\textrm{q}}}}_{\mu\nu}$ to the classical gravitational
field $\tilde{\bar{h^{\textrm{c}}}}_{\mu\nu}$ are given by the formula

\begin{equation}
\tilde{\bar{h^{\textrm{q}}}}_{\mu\nu}(p)=\left(D\Pi\tilde{\bar{h^{\textrm{c}}}}\right)_{\mu\nu}(p),\label{eq:hqprzezhc}
\end{equation}
where
\begin{equation}
D_{\mu\nu}^{\alpha\beta}(p)=\frac{i}{p^{2}}\mathbb{D_{\mu\nu}^{\alpha\beta}}\label{eq:propagator}
\end{equation}
is the free graviton propagator in the harmonic gauge, and $\Pi_{\mu\nu}^{\alpha\beta}(p)$
is the (one-loop) graviton vacuum polarization (self-energy) tensor
operator. Here
\begin{equation}
\mathbb{D}\equiv\mathbb{E}-2\mathbb{P},\quad\textrm{where}\quad\mathbb{E}_{\mu\nu}^{\alpha\beta}\equiv\frac{1}{2}\left(\delta_{\mu}^{\alpha}\delta_{\nu}^{\beta}+\delta_{\nu}^{\alpha}\delta_{\mu}^{\beta}\right),\quad\mathbb{P}_{\mu\nu}^{\alpha\beta}\equiv\frac{1}{4}\eta^{\alpha\beta}\eta_{\mu\nu}.\label{eq:dep}
\end{equation}
Observing that
\begin{equation}
\bar{h}_{\mu\nu}=\left(\mathbb{D}h\right)_{\mu\nu},\label{eq:hbardh}
\end{equation}
we get
\begin{equation}
\tilde{h^{\textrm{q}}}_{\mu\nu}(p)=\frac{i}{p^{2}}\left(\Pi\tilde{\bar{h^{\textrm{c}}}}\right)_{\mu\nu}(p)=\frac{i}{p^{2}}\left(\Pi'\tilde{\bar{h^{\textrm{c}}}}\right)_{\mu\nu}\left(p\right),\label{eq:hqpiprim}
\end{equation}
where the simplified by gauge symmetry version of $\Pi$ equals
\begin{equation}
\Pi'(p)=\kappa^{2}p^{4}I(p^{2})(2\alpha_{1}\mathbb{E}+4\alpha_{2}\mathbb{P}),\label{eq:piprim-ep}
\end{equation}

with the coefficients $\alpha_{1}$ and $\alpha_{2}$ given in Table
\ref{tab:tabela}, and
\begin{equation}
I(p^{2})=-\frac{i}{\left(4\pi\right)^{2}}\log\left(-\frac{p^{2}}{\Lambda^{2}}\right)+\ldots.\label{eq:i-logp2}
\end{equation}
Then the final formula assumes the form:
\begin{equation}
\tilde{h^{\textrm{q}}}_{\mu\nu}(p)=\left(i\kappa^{2}p^{2}I(p^{2})(2\alpha_{1}\mathbb{E}+4\alpha_{2}\mathbb{P})\tilde{\bar{h^{\textrm{c}}}}\right)_{\mu\nu}\left(p\right).\label{eq:hqprzezhcfinal}
\end{equation}

\end{document}